\documentclass[prb,twocolumn,showpacs,amsmath,amssymb,floatfix]{revtex4-1}
\usepackage{graphicx}
\usepackage{color}
\usepackage{pifont}
\usepackage{amstext}
\usepackage{amsmath}
\usepackage{graphics,amssymb}
\usepackage{tabularx}

\begin{document}
\title{Metamagnetism and Lifshitz Transitions in Models for Heavy Fermions.}
\author{M.~Bercx and F.~F.~Assaad}
\affiliation{ Institut f\"ur Theoretische Physik und Astrophysik,
Universit\"at W\"urzburg, Am Hubland, D-97074 W\"urzburg, Germany }
\begin{abstract}
We investigate metamagnetic transitions in models for heavy fermions by considering the doped Kondo lattice model in two dimensions.
Results are obtained within the framework of dynamical mean field and dynamical cluster approximations.
Universal magnetization curves for  different temperatures and Kondo couplings develop upon scaling with the lattice coherence temperature.
Furthermore, the coupling of the local moments to the magnetic field is varied  to take into account the different Land\'e factors of localized and itinerant electrons. 
The competition between  the lattice coherence scale and the  Zeeman energy scale  allows for two  interpretations of the metamagnetism in heavy fermions: 
 Kondo breakdown or Lifshitz transitions.
By tracking the single-particle residue through the  transition,
 we can uniquely conclude  in favor of the Lifshitz transition scenario. 
In this scenario, a quasiparticle band drops below the Fermi energy which leads to a change in topology of the Fermi surface.
\end{abstract}
\maketitle
\section{Introduction}
\label{sec:section1}
Kondo lattice systems are states of matter whose low temperature macroscopic properties are dominated by strong correlations between Bloch fermion states and local spin moments.
They can host various, sometimes competing orders and are therefore susceptible to tuning of external parameters.
How the strongly  entangled Kondo state evolves when competing mechanisms appear, 
constitutes a vibrant area of research.\cite{Gegenwart08, Yamamoto10, Coleman10}
Prototypical heavy fermion materials are intermetallic compounds with the rare-earth elements $\mathrm{Ce}$ or $\mathrm{Yb}$ that deliver almost localized 
$f$-electrons.\\
When an external magnetic field is applied to certain fermion systems, 
unexpected non-linear behavior of the magnetization at a well-defined field value enters the stage. \cite{Haen87,Deppe12}
Equally, distinct anomalies of thermodynamic quantities and in  transport measurements occur at the same magnetic field. \cite{DaouBergemannJulian06,Pfau12}
This phenomenon has been dubbed metamagnetism. 
At the critical field, the heavy electron Fermi surface changes its topology.\cite{Aoki93,DaouBergemannJulian06,Sugi08,Kitagawa11}.
Recent experiments witness a pronounced, first order metamagnetic transition (MMT) in the heavy-fermion paramagnet $\mathrm{CeTiGe}$. \cite{Deppe12}
Metamagnetism has been known  to occur in $\mathrm{CeRu_{2}Si_{2}}$ \cite{Haen87} and, amongst other fermionic systems, a pressure-tuned 
first order MMT has been observed in bilayer ruthenates.\cite{Wu11}\\
The thermodynamic signatures of heavy fermion compounds have been related to a metamagnetic quantum critical endpoint of the Ising universality class. 
\cite{WeickertGarst10,Millis02}
The MMT in heavy fermion systems has been addressed by static mean-field (MF) studies 
\cite{KusminskiyBeachCastroNetoCampbell08, Spalek12} - presupposing a continuous transition - and by dynamical mean field theory (DMFT).
The magnetization profile in Kondo systems has been shown to be closely related to the quasiparticle coherence. \cite{BeachAssaad08}
Also, crystal field effects have been included in a DMFT study. \cite{Yamada11}
Apart from heavy fermion systems, the metamagnetism of itinerant electrons has been addressed by 
MF methods\cite{BinzSigrist04}, functional renormalization group \cite{Honerkamp05} and DMFT. \cite{Bauer09}\\
Lifshitz transitions are quantum phase transitions which invoke a topological change of the Fermi surface.\cite{Lifshitz60,Blanter94,YamajiMisawaImada06} 
Lifshitz transitions and Kondo breakdown scenarios have been investigated in fermionic large-$N$ approaches.\cite{Zhang11, Hackl08}
Zeeman-driven Lifshitz transitions were shown to explain many anomalies in thermodynamic and transport measurements 
of certain heavy fermion metals.\cite{HacklVojta11}\\
This study is motivated by the interplay of two competing energy scales, the lattice coherence scale and a magnetic Zeeman scale.
By varying the magnitude of the Land\'e factors we can show that the metamagnetic transition occurs when both scales are comparable, 
thus allowing for interpretations based on Kondo breakdown or Lifshitz transitions.
The single-particle residue is measured as a function of magnetic field throughout the MMT and is shown to be
consistent with the picture of a coherent band dropping below the Fermi energy at the transition.
We supplement our analysis by single-particle spectral data.
Our results clearly point towards Lifshitz physics as the key player in the MMT in models of heavy fermions.\\
We draw this conclusion  based on a dynamical Cluster Approximation (DCA) calculation of the Kondo lattice model with a Hirsch-Fye quantum Monte Carlo solver.\\
The paper is organized as follows. Section~\ref{sec:section2} introduces the model Hamiltonian and Sec.~\ref{sec:section3} reviews the DCA implementation.
Sections~\ref{sec:section4} and \ref{sec:section5}  contain the results of this study.  We finish with a discussion (Sec.~\ref{sec:section6}) and the conclusion (Sec.~\ref{sec:section7}).

\section{Model}
\label{sec:section2}
The essential aspects of heavy fermion systems  are captured by the Kondo lattice model (KLM).\cite{HewsonBook,Bodensiek11}
The KLM is an effective low-energy model which is obtained upon integrating out the valence fluctuations 
of the $f$-orbitals in the periodic Anderson model.\cite{SchriefferWolff66,TsunetsuguSigristUeda97}
In particular, the model captures the crossover from independent magnetic impurities embedded in a metallic host to a coherent heavy fermion state.
The KLM at half-filling has a unique spin singlet, insulating ground state \cite{Tsunetsugu97} that is adiabatically connected to the trivial band insulator 
of the non-interacting periodic Anderson model.\cite{TsunetsuguSigristUeda97}
The weakly doped KLM exhibits a Fermi liquid ground state.\cite{OtsukiKusunoseKuramoto09}\\
We investigate this model by means of DMFT and DCA
 \cite{HettlerMukherjeeJarrellKrishnamurthy00, JarrelMaierHushcroftMoukouri01}
with a quantum Monte Carlo cluster solver. 
The cluster approximation is on spatial correlations which are essentially cut off by the cluster dimension. 
Temporal correlations that drive the Kondo effect are fully accounted for in DMFT and its cluster extensions.\\
We study the KLM  supplemented with Zeeman terms on the two dimensional square lattice,
 $\mathcal{H} = \mathcal{H}_{\mathrm{t}}  + \mathcal{H}_{\mathrm{J}} + \mathcal{H}_{\mathrm{B}}$:
\begin{eqnarray}\label{eqn:eqn1}
\mathcal{H}&=&
-t\sum_{<i,j>, \sigma}( c^{\dagger}_{i,\sigma}c_{j,\sigma} + \mathrm{H.c.} )
+ J\sum_{i}\mathbf{S}_{i}^{c}\cdot \mathbf{S}_{i}^{f}\nonumber\\
&-& \mu_{B} B \sum_{i} ( g_{c}  S_{z,i}^{c} + g_{f} S_{z,i}^{f} )\;.
\end{eqnarray}
The magnetic moments of itinerant ($c$) and local ($f$) orbitals along the direction of the applied field are given by ${\mu_{z,i}^{c,f} =\mu_{B} g_{c,f} S_{z,i}^{c,f}}$.
The couplings  $g_c$ and $g_f$ are understood as parameters. 
Physically, this is motivated by the pseudo-spin nature of $S_{z,i}^{f}$: 
the spin degree of freedom originates from a Kramer doublet and can take large values which in turn renormalizes the g-factor $g_f$.\cite{Bauer09}\\
Drawing on the recently obtained phase diagram of the two dimensional
 KLM \cite{Watanabe07,OtsukiKusunoseKuramoto09,MartinBercxAssaad10} we concentrate on the paramagnetic side of the transition 
and consider a metallic state with the conduction band filling $n_{c}=0.9$.\\
The lattice of impurities introduces the coherence scale $T_{\mathrm{coh}}$ as the natural energy scale.\cite{Assaad04}
The single-impurity Kondo scale, itself being the natural scale in a single-impurity model, is a local scale. \cite{HewsonBook}\\
Guidelines for these two scales in the KLM are provided by large-$N$ calculations. \cite{BurdinGeorgesGrempel00}
In the weak coupling limit ($J/W \ll 1$, $W$ is the bandwidth) and at small deviation from half-filling ($1-n_{c}\ll 1$)
a scaling of ${T_{\mathrm{coh}}\propto T_{\mathrm{K}}\propto W e^{-\rho_{0}(\epsilon_{\mathrm{F}})/J}}$  is obtained 
($\rho_{0}(\epsilon_{\mathrm{F}})$ is the free density of states at the Fermi level).\\
Two dimensional Kondo systems are realized in surface alloys. e.g. in the heavy fermion compound $\mathrm{CePt_{5}}$. \cite{Klein11, Praetorius12}
In the case that the distance to a continuous quantum critical point is large enough so that the zero dimensional Kondo effect dominates over
 spatial fluctuations one can expect that a similar scenario of competing energy scales applies to the three dimensional case.\\
In the model (Eq.~\ref{eqn:eqn1}) spin-orbit coupling is neglected which would generally cause the g-factor to be a tensor.
Realistic modeling of heavy fermion materials  requires a more sophisticated approach capturing these material specific features. 
Instead, the used model serves the purpose of describing the generic interplay between the magnetic Zeeman scale and the coherence scale of the Kondo system 
which can lead either to the Kondo breakdown or the Lifshitz transition scenario.

\section{Method}
\label{sec:section3}
We use the Hirsch-Fye QMC technique to solve the KLM on  small clusters, that contain two orbitals (DMFT limit) and four orbitals, respectively.
Cluster approximation schemes are particularly well designed to capture the Kondo physics since temporal correlations can be treated exactly. 
The approximation is on spatial correlations that are short ranged in the present situation.
The DCA is a fully causal, non-perturbative method which is systematically  improved by increasing the cluster sizes.
 \cite{HettlerMukherjeeJarrellKrishnamurthy00, JarrelMaierHushcroftMoukouri01}
In the following, we outline our implementation for the KLM. \\
A static mean-field perspective can provide insight into the low energy properties of the KLM. \cite{Burdin09}. 
It roots on the saddle-point approximation which is the exact solution of the $SU(N)$ KLM in the limit of $N=\infty$.
However, it exhibits an unphysical phase transition instead of the Kondo crossover.
Appropriate choices of the magnetic matrix elements that couple the impurity $f$-orbitals to the external magnetic filed can 
recover the smooth Kondo crossover even in the large-$N$ limit of the KLM with an external magnetic field term.\cite{Withoff86}\\
In order to solve the KLM, we implement the following Hamiltonian \cite{BeachAssaad08, CapponiAssaad01}:
\begin{equation}\label{eqn:eqn2}
\mathcal{H} 
=
\mathcal{H}_{\mathrm{0}}
+  \mathcal{H}_{\mathrm{U}}
- \frac{J}{4}\sum_{i}\bigg[ \sum_{\sigma} c^{\dagger}_{i,\sigma}f_{i,\sigma} + f^{\dagger}_{i,\sigma}c_{i,\sigma} \bigg]^{2}\;.
\end{equation}
Here, $\mathcal{H}_{\mathrm{0}} = \mathcal{H}_{\mathrm{t}}  + \mathcal{H}_{\mathrm{B}}$ and the
 Hubbard term $\mathcal{H}_{\mathrm{U}}=\frac{U_{f}}{2}\sum_{i}\big[ \sum_{\sigma} n_{i\sigma}^{f}-1\big]^{2}$ has been introduced. 
Local spin-operators $S_{i}^{f}$ are as usually mapped to auxiliary lattice fermions, 
$S_{i}^{f}=\frac{1}{2}\sum_{\alpha,\beta} f_{i,\alpha}^{\dagger}\mathbf{\sigma}_{\alpha\beta} f_{i,\beta}$. 
Their single occupancy is guaranteed for $U_{f}\rightarrow\infty$ and in this limit, the Hamiltonian (\ref{eqn:eqn2}) is equivalent to the KLM (\ref{eqn:eqn1}).\\
The discretization $\beta=M\Delta\tau$ on the interval $[0,\beta]$ gives the partition function $Z=Z_{\Delta\tau} + \mathcal{O}[(\Delta\tau)^{2}]$, 
with
\begin{eqnarray}\label{eqn:eqn3}
&Z&\!\!_{\Delta\tau}=\mathrm{Tr}\prod_{l=1}^{M}
\bigg\{
\mathrm{exp}\big[-\Delta\tau\mathcal{H}_{\mathrm{0}}\big]\\
&\times&\!\!
\int\!\!\mathrm{D}[\lambda]\;
\mathrm{exp}\bigg[\!\!-i\Delta\tau\sum_{i}\lambda_{li} \big( \sum_{\sigma} n_{i\sigma}^{f}-1 \big) \bigg]\nonumber\\
&\times&\!\!
\int\!\!\mathrm{D}[\phi]\;
\mathrm{exp}\bigg[\!\!-\Delta\tau J\! \sum_{i}\! \bigg(
\phi_{li}^{2} - \phi_{li} \! \sum_{\sigma}\! \big( c^{\dagger}_{i,\sigma}f_{i,\sigma} + \mathrm{H.c.} \big)\! \bigg)\! \bigg]\!
\bigg\}\nonumber\\
&=&\!\!\int\!\!\mathrm{D}[\lambda,\phi]\;
\mathrm{exp}\big[-S_{\mathrm{eff}}[\lambda, \phi]\big]\nonumber\;.
\end{eqnarray}
In Eq.~\ref{eqn:eqn3}, the two successive Hubbard-Stratonovich (HS) transformation reduce the quartic fermion terms to quadratic terms. 
The integration measures  $\mathrm{D}[\lambda,\phi]$ denote integration over  spatial and time indices of the fields and contain normalization factors. \\
The saddle-point of the above defined action fulfills $\partial S_{\mathrm{eff}}/\partial \phi_{li}=\partial S_{\mathrm{eff}}/\partial \lambda_{li}=0$.
Static mean-field theory is obtained by dropping the $\tau$-dependence in the HS fields,
 and one can furthermore request the homogeneous solution:
$\phi_{li}\equiv\phi_{0}\;,\lambda_{li}\equiv\lambda_{0}$.
The saddle-point equations,
\begin{equation}\label{eqn:eqn4}
\phi_{0}=\frac{1}{2N}\langle\sum_{i\sigma} c^{\dagger}_{i,\sigma}f_{i,\sigma} + \mathrm{H.c.} \rangle_{\mathrm{MF}},\;\;
1=\frac{1}{N}\langle\sum_{i\sigma} n_{i,\sigma}^{f} \rangle_{\mathrm{MF}}\;,
\end{equation}
are then solved self-consistently. 
The respective mean field results for total magnetization and quasiparticle residues 
are discussed in Sec.~\ref{sec:section4}.\\
In order to go beyond mean field, a systematic $1/N$ expansion around the mean field solution can be performed. \cite{Saremi07}
Instead, we integrate over all the field configurations through application of the  HF-QMC algorithm.
The trace in Eq.~\ref{eqn:eqn3} can be carried out and expressed as a determinant of the Green's function matrix $g_{\sigma}$.
Then, the partition function
\begin{equation}\label{eqn:eqn5}
Z_{\Delta\tau}
=
\int\mathrm{D}[\lambda,\phi]
\prod_{\sigma}\mathrm{det}\big[g_{\sigma}^{-1}\big]
\end{equation}
is sampled stochastically. 
In the actual implementation, two discrete HS transformation are used. \cite{CapponiAssaad01}
The Green's function matrix is measured and updated according to the Hirsch-Fye algorithm.\cite{AssaadBook08}
During the simulation of Eq.~\ref{eqn:eqn2},  double occupancy of the $f$-orbitals can be suppressed to the desired accuracy.
We take $\Delta\tau=0.25$ during the simulations. 
We have checked that smaller values of $\Delta\tau$ do not alter the results.\\
The cluster approximation amounts to considering the interaction terms of the Hamiltonian only on a subset $\mathcal{M}$ of the lattice with $N_{c}$ sites,
 which naturally defines the extent to which spatial correlations are captured.
We therefore solve the model
\begin{equation}\label{eqn:eqn6}
\mathcal{H}=\tilde{\mathcal{H}}_{\mathrm{0}} + J\sum_{R\in\mathcal{M}}\mathbf{S}_{R}^{c}\cdot \mathbf{S}_{R}^{f}\;,
\end{equation}
by using the auxiliary Hamiltonian (\ref{eqn:eqn2}).
$\tilde{\mathcal{H}}_{\mathrm{0}}$ denotes the bath which is determined self-consistently.\\
The DCA is naturally described in momentum space since it relies on  coarse-graining of momentum space.
Since the interaction  part of the Hamiltonian is local, it is not affected by the coarse-graining.
The model Hamiltonian is solved on a finite cluster of $N_{c}$ sites that is embedded in a bath of $N$ sites ($N\gg N_{c}$).
 Since $N$ is not a limiting factor one can work directly in the thermodynamic limit.
Therefore, the DCA interpolates between two limiting cases: the DMFT ($N_{c}=1$) and the finite lattice ($N_{c}=N$).\\
The DCA lattice self energy is a step-function in reciprocal space:
\begin{eqnarray}\label{eqn:eqn7}
\Sigma^{\mathrm{DCA}}_{\mathrm{Latt}}(\mathbf{K},\omega) 
&=& 
\frac{N_{c}}{N}\sum\limits_{\tilde{\mathbf{k}}}\Sigma_{\mathrm{Latt}}(\mathbf{K} +\tilde{\mathbf{k}},\omega)\;,\\
\underset{N_{c}\rightarrow N}{\mathrm{lim}} \Sigma^{\mathrm{DCA}}_{\mathrm{Latt}}(\mathbf{K},\omega) &=& \Sigma_{\mathrm{Latt}}(\mathbf{K},\omega)\;.\nonumber
\end{eqnarray}
The step size is $\Delta \mathbf{K}=2\pi/N_{c}$, the cluster momenta $\mathbf{K}$ define the centers of $N_{c}$ reciprocal cells 
and $\tilde{\mathbf{k}}$ denotes the k-points that lie within these cells.
The DCA self-consistent scheme operates on the single-particle level of the self energies and it demands that
 $\Sigma^{\mathrm{DCA}} _{\mathrm{Cluster}}(\mathbf{K},\omega) =\Sigma^{\mathrm{DCA}}_{\mathrm{Latt}}(\mathbf{K},\omega)$.
The self-consistent equations  are:
\begin{eqnarray}\label{eqn:eqn8}
&\Sigma&^{\mathrm{DCA}}_{\mathrm{Cluster}}(\mathbf{K},\omega)\\
&=&
G^{\mathrm{DCA}}_{\mathrm{Latt,av.}}(\mathbf{K},\omega)^{-1} + \Sigma^{\mathrm{DCA}}_{\mathrm{Latt}} (\mathbf{K},\omega) - 
 G^{\mathrm{DCA}}_{\mathrm{Cluster}}(\mathbf{K},\omega)^{-1}\nonumber
\end{eqnarray}
Here, an effective bare Green function  has been defined as 
\begin{equation}\label{eqn:eqn9}
\mathcal{G}^{\mathrm{DCA}}_{\mathrm{Cluster}}(\mathbf{K},\omega) ^{-1} 
= 
G^{\mathrm{DCA}}_{\mathrm{Latt, av.}}(\mathbf{K},\omega)^{-1} + \Sigma^{\mathrm{DCA}}_{\mathrm{Latt}} (\mathbf{K},\omega) \;.
\end{equation}
The DCA lattice averaged Green functions are coarse-grained averages over cell momenta:
\begin{equation}\label{eqn:eqn10}
G^{\mathrm{DCA}}_{\mathrm{Latt, av.}}(\mathbf{K},\omega)
=
\frac{N_{c}}{N}\sum\limits_{\tilde{\mathbf{k}}}\frac{1}{\omega-\epsilon(\mathbf{K}+\tilde{\mathbf{k}})+\mu 
- \Sigma^{\mathrm{DCA}}_{\mathrm{Latt}}(\mathbf{K},\omega)} \;.
\end{equation}
The Green function $\mathcal{G}^{\mathrm{DCA}}_{\mathrm{Cluster}}(\mathbf{K},\omega) $ is the bare Green function that is the input for the cluster calculation.
The cluster calculation yields the cluster Green functions $G^{\mathrm{DCA}}_{\mathrm{Cluster}}(\mathbf{K},\omega)$ that enter Eq.~\ref{eqn:eqn8}.
Once the self-energy is converged, the DCA lattice Green function is computed: 
\begin{equation}\label{eqn:eqn11}
G^{\mathrm{DCA}}_{\mathrm{Latt}}(\mathbf{k},\omega)^{-1}
=\omega-\epsilon(\mathbf{k})+\mu - \Sigma^{\mathrm{DCA}}_{\mathrm{Latt}}(M(\mathbf{k}),\omega)\;. 
\end{equation}
The function $M:\mathbf{k}\rightarrow \mathbf{K}$ uniquely maps momenta to the reciprocal cells.\\
The required  CPU time of the HF-QMC algorithm scales as $(\beta N_{c})^{3}$.

\section{Results}
\label{sec:section4}
The magnetic field tunes  the interacting Kondo system (Eq.~\ref{eqn:eqn1}) from strong coupling at low fields to weak coupling at high fields, $B/T_{\mathrm{coh}} \gg 1$. 
This limit is adiabatically connected to two copies of non-interacting $c$-electrons, spin split by the Zeeman energy, and fully polarized $f$-moments.
At low values of the magnetic field, the hybridized band is expected to shift in a rigid manner.
At an intermediate energy scale, $B \sim T_{\mathrm{coh}}$, two different scenarios are conceivable: 
(1), a breakdown of the Kondo effect itself at the relevant energy scale or, 
(2), a continuous transition that preserves the quasiparticles.
In scenario (1), the quasiparticle itself is destroyed by the magnetic field.
The single-particle residue quantifies the overlap of the interacting wave function with a bare conduction electron wave function.
Therefore, the loss of quasiparticle coherence has to manifest itself as a sudden drop in this quantity for both both spin projections.
In scenario (2), quasiparticles remain intact at the Fermi level.
The spin dependent Fermi surfaces undergo Lifshitz transitions which modify their topology.
As shown below, data for the single-particle residue and single-particle spectral function across the MMT support scenario (2).\\
The mean-field solution, derived from the saddle-point of the $SU(N)$ KLM,
 can be seen as the best approximation in quadratic fermionic terms to the fully correlated model.
Therefore, in the case that DMFT/DCA calculations support the notion of quasiparticles, the MF perspective is legitimate.
The MF results are intended to complete the above described scenario of Lifshitz transitions.\\
The magnetization profile of a heavy fermion model system can serve directly as a measure of coherence.
The plateau of the occupation number difference, 
${m=\sum_{\sigma}\sigma\big( n_{\sigma}^{c} +  n_{\sigma}^{f}\big)}$ directly relates to the hybridization gap in the quasiparticle bands.\cite{BeachAssaad08} 
Its position is fixed to $x=1-n_{c}$ by the  Luttinger sum rule.
The physical magnetization  ${M=-\frac{\partial \mathcal{F}}{\partial B}=  \sum_{\sigma}\sigma\big( g_{c} n_{\sigma}^{c} +  g_{f} n_{\sigma}^{f}\big)}$ 
does not generally display a plateau when the orbital couplings are not the same.

\subsection{Data collapse $M(g_{f}/g_{c},T/T_{\mathrm{coh}}, B/T_{\mathrm{coh}})$ }
\begin{figure}
  \begin{center}
  \includegraphics[width=\columnwidth,type=png,ext=.png,read=.png]{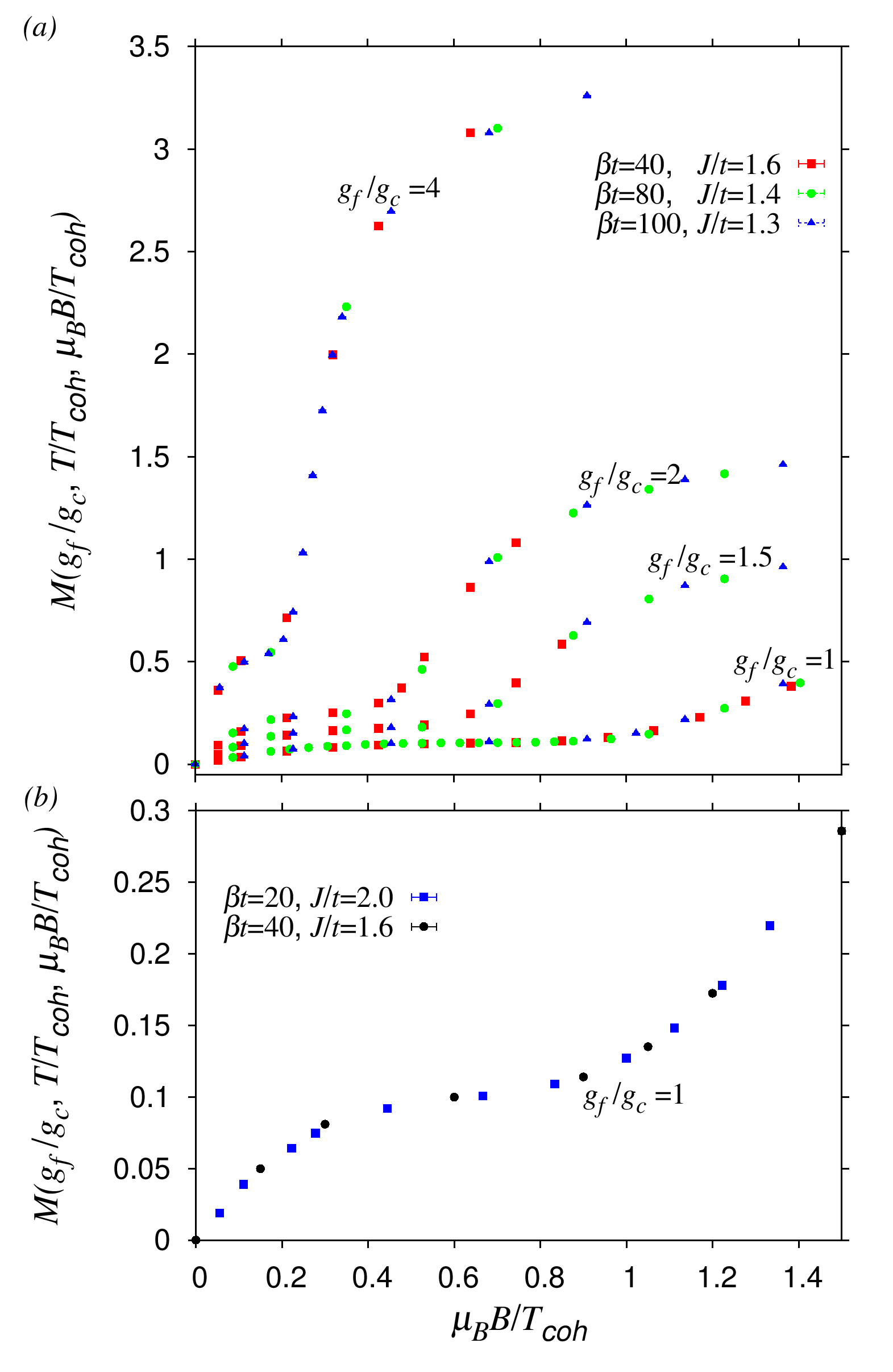}
  \end{center}
  \caption{(Color online) $T_{\mathrm{coh}}$-scaled magnetization $M(g_{f}/g_{c},T/T_{\mathrm{coh}}, B/T_{\mathrm{coh}})$ for (a) two-orbital DMFT  and (b) four-orbital DCA.\\
The respective coherence temperatures are:
(a)~
$T_{\mathrm{coh}} (g_{f}/g_{c}=1, J/t=1.6) =0.094 t$, $T_{\mathrm{coh}} (g_{f}/g_{c}=1, J/t=1.4) =0.057 t$ and
$T_{\mathrm{coh}} (g_{f}/g_{c}=1, J/t=1.3) =0.044 t$; 
(b)~$T_{\mathrm{coh}} (g_{f}/g_{c}=1, J/t=1.6) =0.10 t$ and $T_{\mathrm{coh}} (g_{f}/g_{c}=1, J/t=2.0) =0.18 t$.
}
  \label{fig:fig1}
\end{figure}

\begin{figure}
  \begin{center}
  \includegraphics[width=\columnwidth,type=png,ext=.png,read=.png]{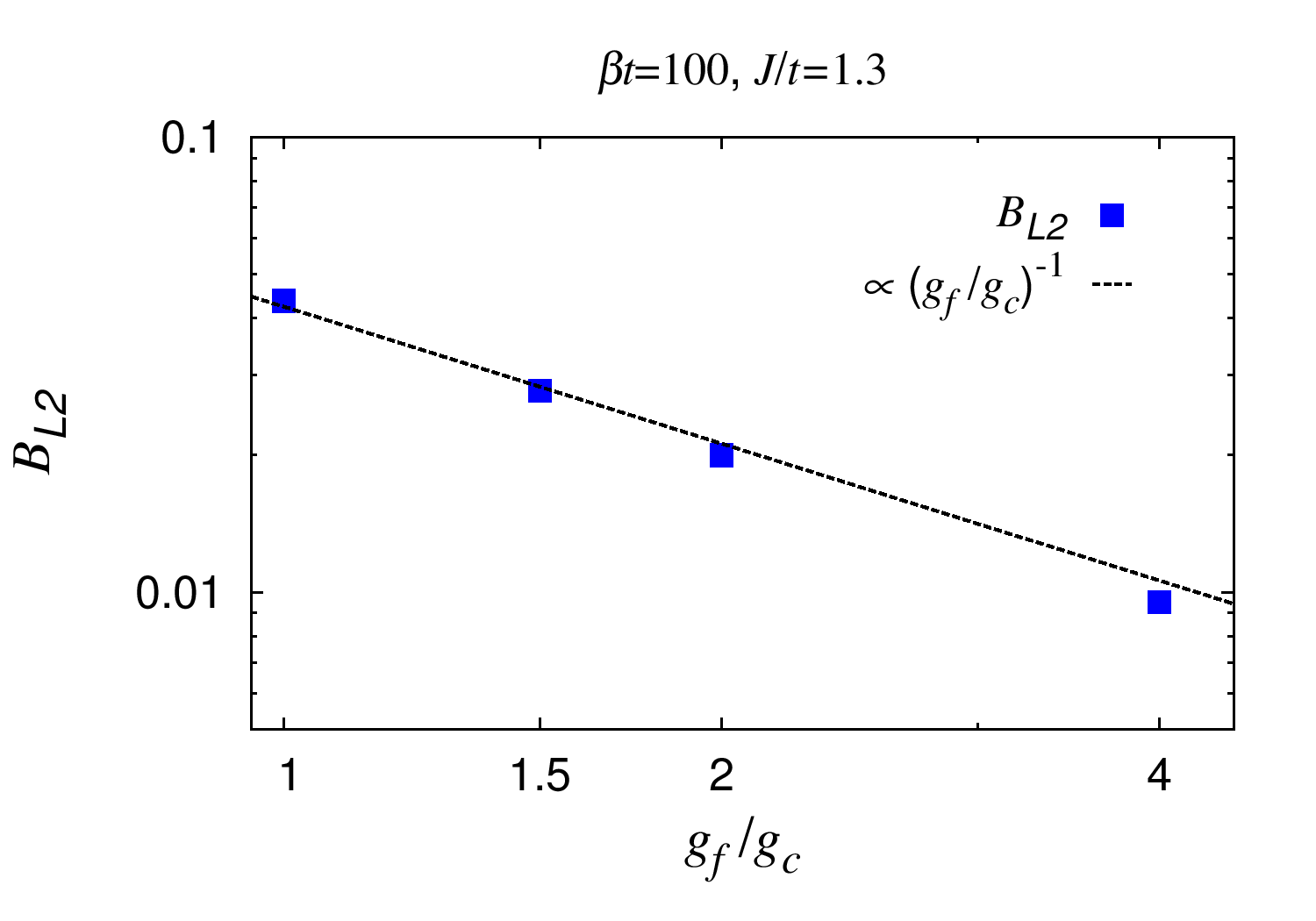}
  \end{center}
  \caption{(Color online) The magnetic field value of the second Lifshitz transition, $B_{\mathrm{L2}}(g_{f}/g_{c})$, 
 agrees well with  $B_{\mathrm{L2}}(g_{f}/g_{c})\propto\big( g_{f}/g_{c} \big)^{-1}$.
}
  \label{fig:fig2}
\end{figure}

\begin{figure}
  \begin{center}
  \includegraphics[width=\columnwidth,type=png,ext=.png,read=.png]{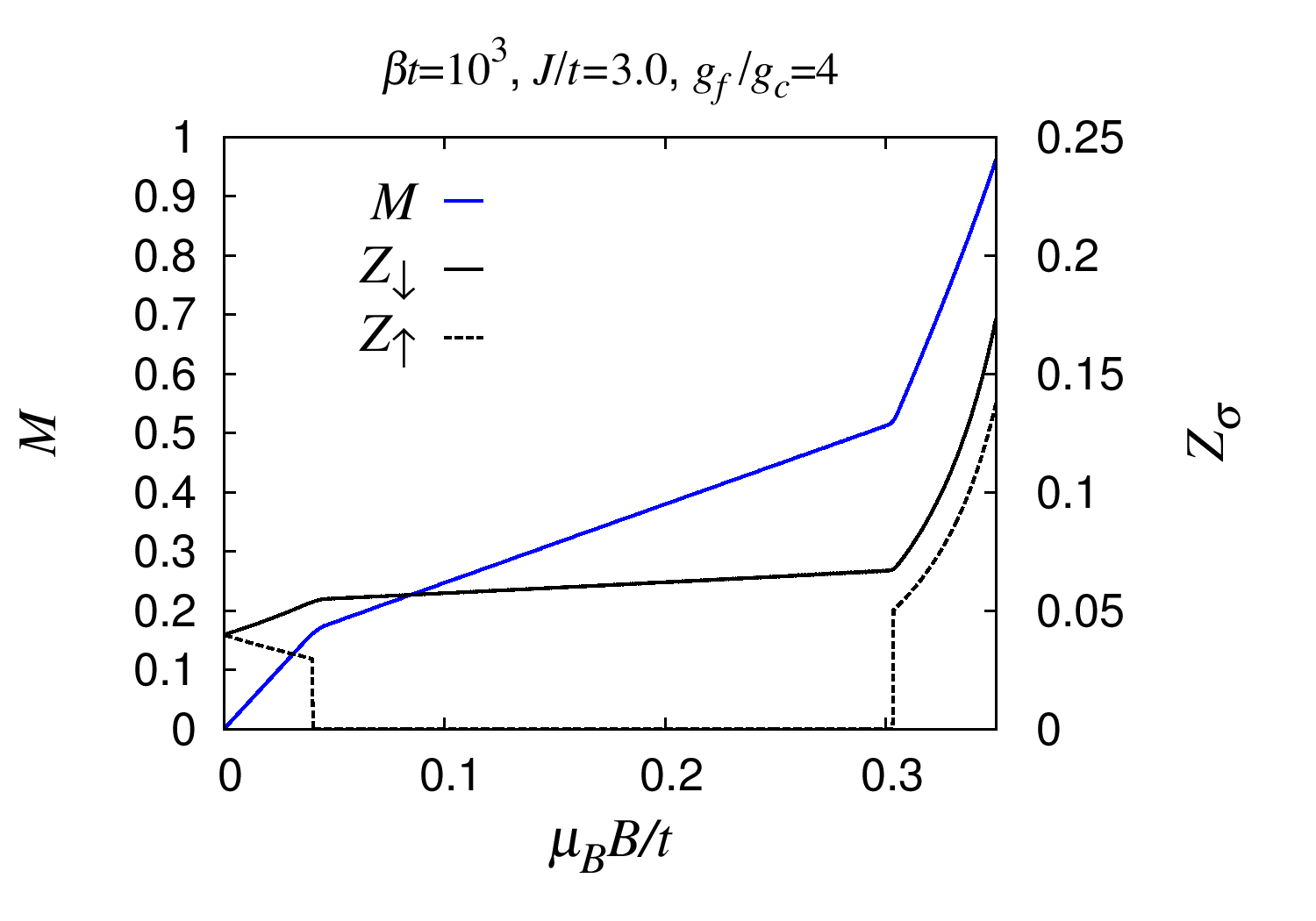}
  \end{center}
  \caption{(Color online)  Static MF results for the magnetization $M$ and single-particle residue $Z_{\sigma}$.}
  \label{fig:fig3}
\end{figure}

For temperatures below the coherence scale $T_{\mathrm{coh}}$ quasiparticle bands are formed via coherent superposition of the screening clouds of the local spins.
To verify that the coherence scale is the unique  underlying scale a data collapse of DMFT and DCA data is carried out 
by scaling the magnetization with $T_{\mathrm{coh}}$.
Because the plateau width of the occupation number is a measure of the hybridization gap, 
a good estimate of $T_{\mathrm{coh}}$ is obtained from the position of the second kink at $B=B_{\mathrm{L2}}$,
determined by the intersections of linear fits at $\beta t=100$ and $g_{f}/g_{c}=1$.  
The scaling then becomes:
\begin{equation}\label{eqn:eqn12}
M(g_{f}/g_{c},T,B,J)
 \rightarrow M(g_{f}/g_{c},T/T_{\mathrm{coh}}, B/T_{\mathrm{coh}})\;. 
\end{equation}
Effectively, the Kondo coupling $J/t$ has disappeared as a parameter in $M$.
The data collapse is evident, as shown in Fig.~\ref{fig:fig1}.\\
For all values of $g_{f}/g_{c}$ the magnetization shows two pronounced kinks at $B=B_{\mathrm{L1,2}}$.
The driving mechanism that shapes the magnetization is rooted in the competition of two 
energy scales: the dominant magnetic energy scale $g_{f}\mu_{B}B$ and the Kondo scale $T_{\mathrm{coh}}$.
At the second kink both scales become comparable, such that $B_{\mathrm{L2}}\propto g_{f}^{-1}$.
The position of the second kink in dependence of the coupling ratio is shown in Fig.~\ref{fig:fig2}
and the data are in good agreement with the above argument.
Increased Zeeman coupling to the local spins provokes the  intermediate, plateau-like region to decrease and 
renders the increase at  $B=B_{\mathrm{L2}}$ much steeper.\\
Static MF calculations succeed in reproducing the qualitative shape of $M$ (Fig.~\ref{fig:fig3}). 
In the MF picture, the two kinks in the magnetization correspond to two Lifshitz transitions.\\
At this point, the data collapse of the magnetization can be compared to a scaling approach of the resistivity in a recent cluster DMFT (CDMFT) study of the Anderson lattice model 
close to half-filling of the conduction band, which equally reveals the lattice coherence temperature as the single underlying energy scale. \cite{Tanaskovic11}\\
Our calculated metamagnetic curves, as shown in Fig.~\ref{fig:fig1},  
bear notable similarity with recent experimental data of the paramagnetic heavy-fermion system $\mathrm{CeTiGe}$. \cite{Deppe12}
This is discussed in Sec. \ref{sec:section6}.

\subsection{Single-particle quantities: residue $Z_\sigma$ and spectral function $A_{\sigma}(\mathbf{k},\omega)$ }
\begin{figure}
  \begin{center}
  \includegraphics[width=\columnwidth,type=png,ext=.png,read=.png]{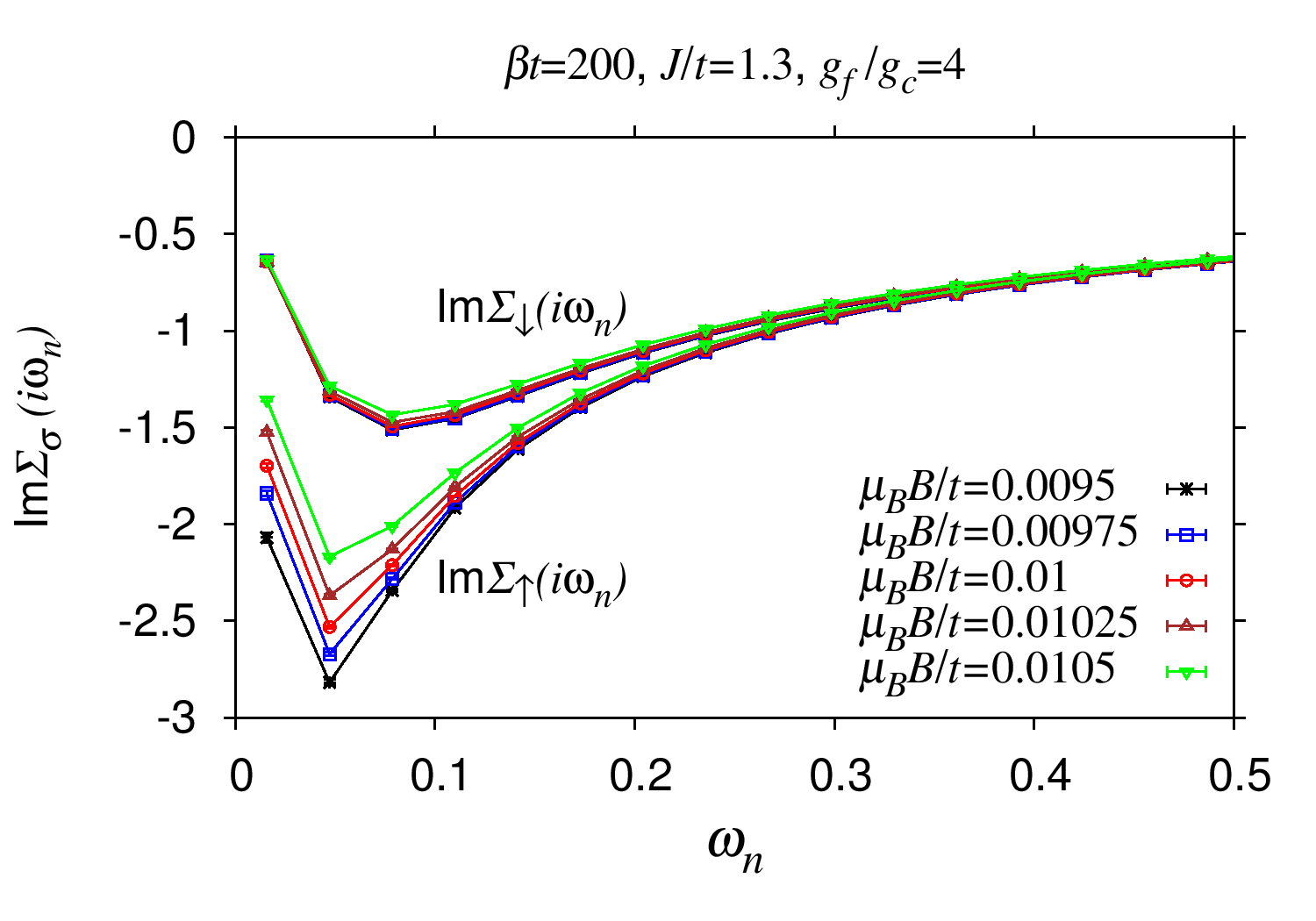}
  \end{center}
  \caption{(Color online) The imaginary part of Matsubara self energies
 $\mathrm{Im}\Sigma^{\mathrm{DMFT}}_{\sigma=\downarrow,\uparrow}(i\omega_{n})$ at values of the magnetic field  close to ${B=B_{\mathrm{L2}}}$. }
  \label{fig:fig4}
\end{figure}

\begin{figure}
  \begin{center}
  \includegraphics[width=\columnwidth,type=png,ext=.png,read=.png]{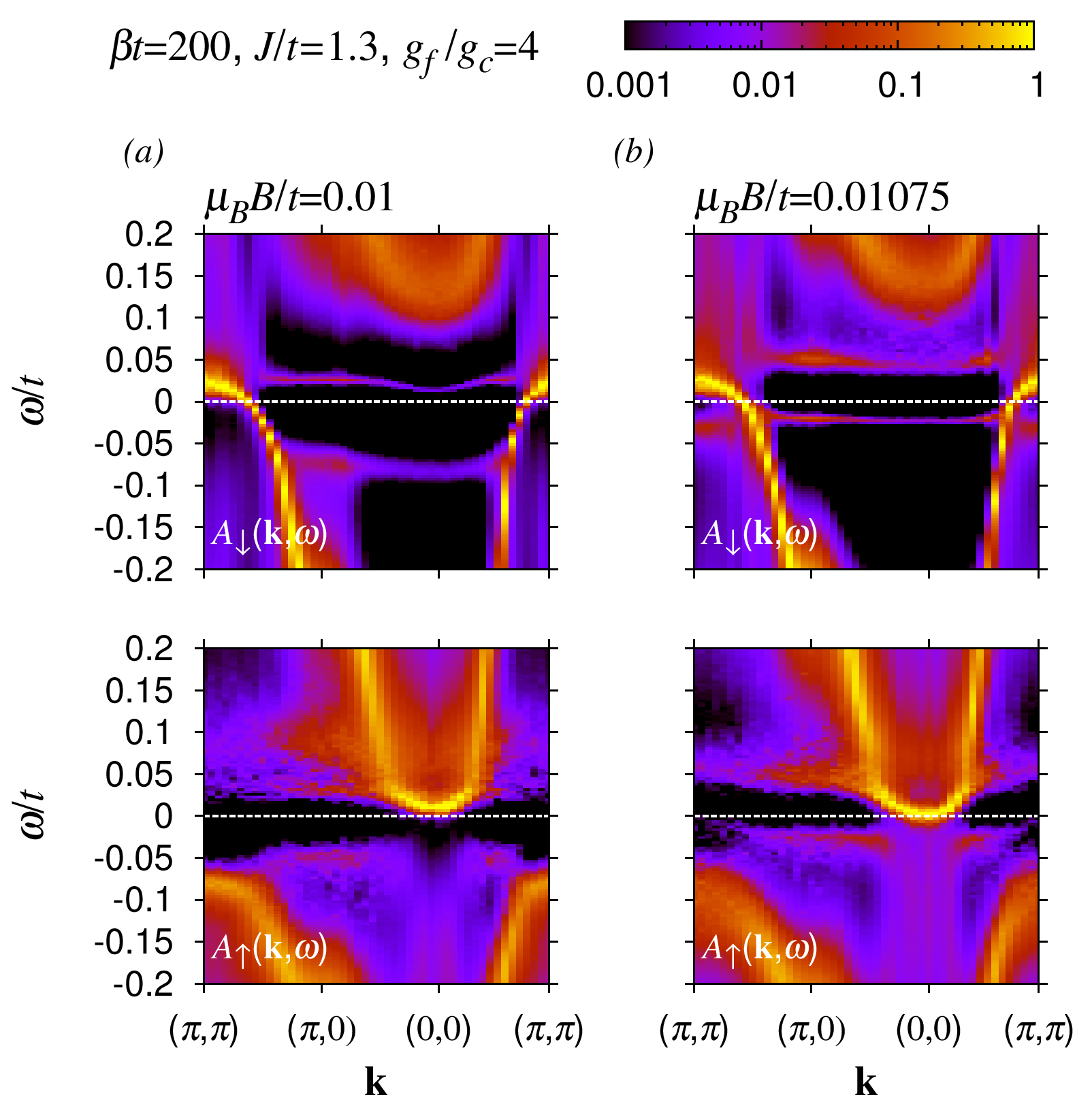}
  \end{center}
  \caption{(Color online) Single-particle spectral function $A_{\sigma}(\mathbf{k},\omega)$ across the MMT at the lowest temperature, $\beta t=200$ 
(magnetic field values are indicated by arrows in Fig.~\ref{fig:fig6} ).
The narrow distribution of spectral weight close to the Fermi energy (dashed line) indicates that Kondo coherence remains across the MMT.}
  \label{fig:fig5}
\end{figure}

\begin{figure}
  \begin{center}
  \includegraphics[width=\columnwidth,type=png,ext=.png,read=.png]{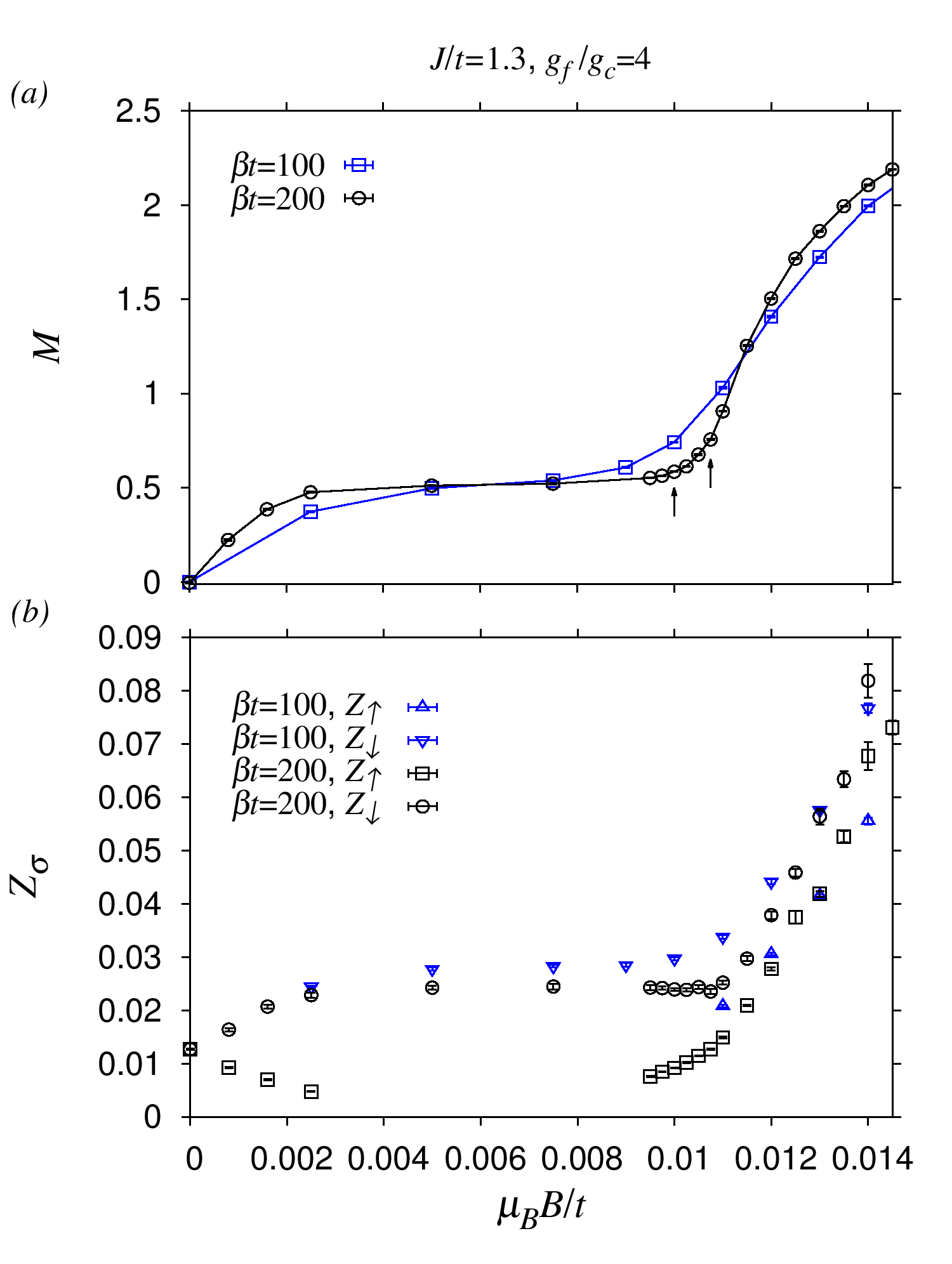}
  \end{center}
  \caption{(Color online) Magnetization $M$ and single-particle residue $Z_{\sigma}^{\mathrm{DMFT}}$ across the MMT. 
(The arrows refer to the single-particle spectra of Fig.~\ref{fig:fig5}.)}
  \label{fig:fig6}
\end{figure}

The analysis of the single-particle quantities is based on the observation,
 that the KLM has a Fermi liquid ground state for the chosen value of conduction band filling, $n_{\mathrm{c}}=0.9$.\cite{OtsukiKusunoseKuramoto09}, 
and for zero external magnetic field.
The calculations were performed for $J/t = 1.3$, $g_{f}/g_{c}=4$, $\beta t=100$ and $\beta t=200$. 
For these parameters, we identify two Lifshitz transitions that occur at $\mu_{\mathrm{B}}B_{\mathrm{L1}}/t\approx0.002$ and
at $\mu_{\mathrm{B}}B_{\mathrm{L2}}/t\approx0.01075$. The latter corresponds to the MMT.\\
A Fermi liquid signature is the analyticity of the retarded self energy $\Sigma(\mathbf{k},\omega)$ 
around the Fermi energy such that $\Sigma(\mathbf{k},\omega)$ allows for polynomial expansion. 
Then, the single-particle residue reads, expressed with the k-independent
Matsubara self-energy $\Sigma_{\sigma}^{\mathrm{DMFT}}(i\omega_{n})$:
\begin{equation}\label{eqn:eqn13}
\Big[Z_{\sigma}^{\mathrm{DMFT}}\Big]^{-1}
=\underset{T\rightarrow 0}{\mathrm{lim}} 
\bigg[1-\frac{\mathrm{Im}\Sigma_{\sigma}^{\mathrm{DMFT}}(\omega_{n})}{\omega_{n}}\bigg]_{\omega_{n}=\pi T}\;.
\end{equation} 
The quantity $\mathrm{Im}\Sigma_{\sigma}^{\mathrm{DMFT}}(\omega_{n})$ across the MMT at $B=B_{\mathrm{L2}}$ is displayed in Fig.~\ref{fig:fig4}.
Evidently, the imaginary part of the Matsubara self energy is free of divergences for both spin projections at low frequencies $\omega_{n}$ .
We take this as evidence for the continuous transition scenario.\\
The excitations are tracked by  the single-particle spectral function,
\begin{equation}\label{eqn:eqn14}
A_{\sigma}(\mathbf{k},\omega) = -\frac{1}{\pi}\mathrm{Im}G_{\mathrm{Latt}}^{\sigma}(\mathbf{k},\omega)\;, 
\end{equation}
see Fig.~\ref{fig:fig5}.
The analytic continuation from imaginary time dependent QMC data has been performed with the stochastic maximum entropy method.\cite{Beach04}\\
The single-particle residues  $Z_{\sigma}$ at $\beta t=100$ and $\beta t=200$ across the MMT are shown in Fig.~\ref{fig:fig6}(b).
$Z_{\downarrow}$ essentially follows the magnetization  $M(B)$ (Fig.~\ref{fig:fig6}(a)). 
$A_{\downarrow}(\mathbf{k},\omega)$ displays well defined quasiparticle weight across the MMT (Fig.~\ref{fig:fig5}) 
and hence accounts for a metallic state.\\
$Z_{\uparrow}$ vanishes for  an intermediate magnetic field range, close to $B_{L1}<B<B_{L2}$.
In this locked phase, no up-spin Fermi surface is present.
At  $B=B_{\mathrm{L2}}$, a topological change of the Fermi surface occurs since one up-spin band crosses the Fermi level at the gamma point, $(k_{x}, k_{y})=(0,0)$.
$A_{\uparrow}(\mathbf{k},\omega)$ shows a sharply defined quasiparticle band just below and at $B=B_{\mathrm{L2}}$, see Fig.~\ref{fig:fig5} (a) and (b).
The fact that the residue $Z_{\uparrow}$ does not vanish exactly at  $B=B_{\mathrm{L2}}$  can be related to the finite temperature.
Also, we note that the single-particle residue is not fully converged in the intermediate field range, even at the lowest temperatures.\\
In the static MF scenario, the two Lifshitz transitions are naturally present.
As shown in Fig.~\ref{fig:fig3}, the single-particle residue $Z_{\uparrow}$, calculated from the MF coherence factors at the Fermi energy, 
displays the expected step-like behavior.\\
The Lifshitz transition at $B=B_{\mathrm{L2}}$ equally marks the transition from heavy to light fermions 
which is reflected in the  steep increase of   $Z_{\sigma}$ as the magnetic field is ramped up further, see Fig.~\ref{fig:fig6}(b).
This is in accordance with the notion of adiabatic continuity to free fermions 
which is expected in the limit of high magnetic fields, i.e. weak coupling. \cite{BenlagraPruschkeVojta11}
Based on the $\beta t=200$ DMFT results, we conclude that a continuous transition from low to high magnetic fields occurs, 
at least at and above this temperature.

\section{Beyond DMFT}
\label{sec:section5}
\begin{figure}
  \begin{center}
  \includegraphics[width=\columnwidth,type=png,ext=.png,read=.png]{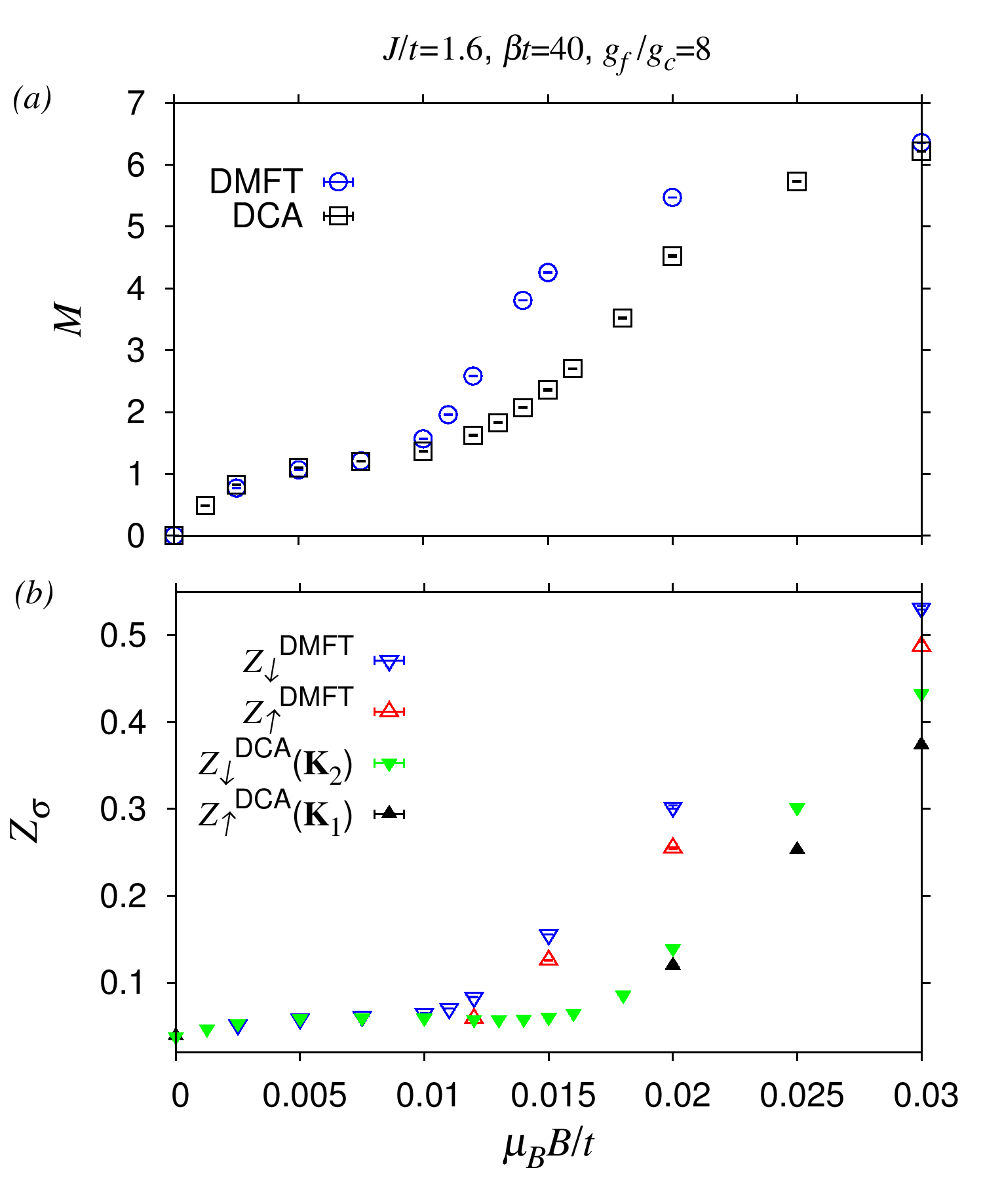}
  \end{center}
  \caption{(Color online) Magnetization $M$ and single-particle residues $Z_{\sigma}$ from two-orbital (DMFT) calculations and four-orbital (DCA) calculations. 
The $k$-vectors $\mathbf{K}_{1}=(0,0)$ and $\mathbf{K}_{2}=(\pi,\pi)$ denote the relevant DCA patches.}
  \label{fig:fig7}
\end{figure}

The DCA calculates the k-dependent self-energy $\Sigma_{\sigma}^{\mathrm{DCA}}(\omega_{n}, \mathbf{K})$. 
This leads to the estimate for the residue 
\begin{eqnarray}\label{eqn:eqn15}
&\Big[&Z^{\mathrm{DCA}}_{\sigma}(M(\mathbf{k}_{f}))\Big]^{-1}\\
&=&\underset{T\rightarrow 0}{\mathrm{lim}} 
\bigg[1-\frac{ \mathrm{Im}\Sigma_{\sigma}^{\mathrm{DCA}}(\omega_{n}, M(\mathbf{k}_{f}))}{\omega_{n}}\bigg]_{\omega_{n}=\pi T}\nonumber\;,
\end{eqnarray}
The map function $M:\mathbf{k}_{f}\rightarrow \mathbf{K}$ maps the Fermi momentum to the matching reciprocal patch.\\
The 4-orbital DCA measurements agree with the 2-orbital DMFT results 
in the limits of strong coupling (small magnetic field) and weak coupling (large magnetic fields), see Fig.~\ref{fig:fig7}(a).
In the intermediate regime, around $B=B_{\mathrm{L2}}$, deviations are detected in the magnetization as well as in the 
single-particle residues. 
The inclusion of spatial fluctuations softens the transition considerably.
This  can be understood from the notion of  an effective Land\'e-factor $g_{f}$ which becomes lower when spatial fluctuations are present, since,
on the two-site cluster, the local moment can be quenched not only dynamically but also via local singlet formation.
The single-particle residue in the down-spin projection displays no sign of vanishing across the MMT (Fig.~\ref{fig:fig7}(b)).

\section{Discussion}
\label{sec:section6}
Lifshitz transitions are continuous quantum phase transition which do not change symmetry but Fermi surface topology. \cite{Lifshitz60}
Strictly speaking, they are defined for free fermion systems at zero temperature.
Due to the unambiguous presence of quasiparticles, the notion of Lifshitz transition can be carried over to the KLM.
Driven by the external magnetic field, two consecutive Lifshitz transitions take place, at  $B=B_{\mathrm{L1,2}}$, 
and the second one is identified with the MMT.
This scenario is maintained when the $f$-moments are allowed to couple more strongly to the field by altering the ratio $g_{f}/g_{c}$.\\
Collective effects challenging the quasiparticle coherence seem to be of minor importance during the MMT, even when $B\sim T_{\mathrm{coh}}$.
Naturally, our calculation scheme is limited to the dominantly paramagnetic regime of the KLM.
The choice of parameters, $n_{c}=0.9$ and $J/t\geq1.3$  place our results unambiguously in the paramagnetic phase. 
\cite{Watanabe07,OtsukiKusunoseKuramoto09,MartinBercxAssaad10}
First steps (Fig.~\ref{fig:fig7}) in a systematic DCA study of larger clusters
that can  take into account the Ruderman-Kittel-Kasuya-Yosida (RKKY) interaction between local moments leave the Lifshitz scenario at the MMT invariant.
This is consistent with the fact that temporal fluctuations that generate the Kondo effect 
dominate the physics at the MMT.
Close to a critical point where the range of spatial fluctuations becomes large
our approximation will fail and another modeling will be required.\\
Transport  signatures of the Lifshitz transition can be calculated  with the Boltzmann transport approximation.
Topological changes of the bands that cross the Fermi energy can strongly influence transport measurements, in particular when these bands are shallow.
This offers an explanation for the anomalies observed in Zeeman driven heavy-fermion systems.\cite{HacklVojta11}\\
Compared to our results for the magnetic field dependent single-particle spectrum (Fig.~\ref{fig:fig5}), similar results have been obtained
for  the ferromagnetic phase of the Kondo lattice model without external field terms. \cite{Peters12}
There, the spin-dependent shift of the quasiparticle weight is generated dynamically and leads to the notion of a spin-selective Kondo insulating phase.\\
Our results are applicable to heavy fermion compounds that  have a magnetic field-driven Lifshitz transition at the coherence scale.\\
The materials $\mathrm{CeTiGe}$ \cite{Deppe12} and $\mathrm{CeRu_{2}Si_{2}}$ \cite{DaouBergemannJulian06, Flouquet02} have a MMT at 
magnetic energy scales that are consistent with their estimated coherence temperatures.
In our model, the critical metamagnetic field corresponds to the second Lifshitz transition at $B_{\mathrm{L2}}$.
In this mechanism of competing energy scales we expect that the details of the band structure are of secondary importance. 
This is in contrast to Lifshitz transitions at magnetic fields much below the coherence scale where 
the details of the band structure are essential. \cite{HacklVojta11}\\
The metamagnetic signatures of our model (Fig.~\ref{fig:fig1}) are  similar to recent experimental data of the  
paramagnetic 4$f$-based compound $\mathrm{CeTiGe}$  which exhibits a pronounced first-order MMT. \cite{Deppe12}
Its anticipated coherence scale,  $T_{\mathrm{coh}}\approx 55\;\mathrm{K}$, 
is of the same order as the critical magnetic field of $\mu_{0}B_{\mathrm{MMT}}=12.5\;\mathrm{T}$, assuming in our model a g-factor $g_{f}\approx 7$.
Equally, at lower fields, the magnetization is found to slightly change its slope, 
which might correspond to a first Lifshitz transition which in our model happens at $B_{\mathrm{L1}}$.
The experimentally observed distinct drop of the effective quasiparticle mass  is in accordance with our findings for the KLM (see Sec. \ref{sec:section4}).
Importantly, we find the MMT to be continuous both in the two-orbital DMFT and in the four-orbital DCA calculations
 and on the temperature scales we can access.\\
$\mathrm{CeRu_{2}Si_{2}}$ exhibits a continuous MMT and simultaneously a Zeeman-driven topology change of the Fermi surface \cite{DaouBergemannJulian06, Pfau12}.  
The magnetization increases seemingly linear as the magnetic field is increased towards the metamagnetic field. \cite{Flouquet02} 
The critical field  $\mu_{0}B_{\mathrm{MMT}}=7.8\;\mathrm{T}$ matches the  coherence temperature of $T_{\mathrm{coh}}\approx 20\;\mathrm{K}$ \cite{DaouBergemannJulian06, Flouquet02} 
when the g-factor in our model is assumed to be  $g_{f}\approx 4$. 
A Lifshitz transition at the coherence scale is therefore a plausible scenario for the MMT in $\mathrm{CeRu_{2}Si_{2}}$.

\section{Conclusion}
\label{sec:section7}
We have explored the Zeeman driven MMT in the Kondo lattice model
which is considered to be the paradigmatic low energy model for heavy fermion systems.
Results for the paramagnetic metallic phase of the KLM are obtained in the framework of DMFT/DCA which can exactly account for the Kondo effect.\\
Upon scaling the relevant energy scales with the lattice coherence scale  the collapse of the magnetization data to a universal curve is observed,
 independent of the Kondo interaction.
This data collapse has been confirmed for a range of Kondo couplings, temperatures and ratios of Land\'e factors.
The pseudo spin nature of the $f$-orbitals, resulting from a Kramer's doublet, can be taken in account with an effective Land\'e factor $g_{f}$
and the competition of magnetic scale and coherence scale is invariant on the choice of $g_{f}$.\\
We have traced the single-particle residue  from low to high magnetic fields and 
report that it is continuous at the lowest temperatures our simulation can access.
Two consecutive Lifshitz transitions occur as the field is ramped up and cause the change in topology of the spin-projected Fermi surfaces.
This lead us to the finding that the MMT in the KLM  is coincident with a continuous Lifshitz transition.
The absence of a singularity in the single-particle residue at the MMT excludes the Kondo breakdown scenario.\\
At the temperature scale we can access, the sharp increase of magnetization at the MMT can  well be
explained as a consequence of a continuous Lifshitz transition in heavy fermion model systems 
where the Land\'e factor of the local spins is larger than the one for the itinerant electrons.
In the course of this transition the excitations change their character from heavy fermions to light fermions.\\
The recently observed first order nature \cite{Deppe12} of the metamagnetic phase transition at a temperature $T\ll T_{\mathrm{coh}}$ remains an open issue. 
Of particular importance is understanding if the KLM itself can account for the low temperature first order nature of the transition 
or if other competing energy scales such as coupling to the lattice \cite{Raczkowski10} have to be taken into consideration.

\begin{acknowledgments}
We would like to thank A.~Benlagra, M.~Brando and H.~Pfau for discussions and F.~Goth for careful reading of the manuscript.
We acknowledge support from the DFG under Grant No.~FOR1162. 
The numerical simulations were carried out at the J\"ulich Supercomputing Centre 
and we thank this institution for generous allocation of CPU time.
\end{acknowledgments}
\bibliography{bercx2012.bib}
\end{document}